\documentclass[aps,twocolumn,showpacs,amssymb]{revtex4-1}

\usepackage[dvips]{graphicx}% Include figure files
\usepackage{dcolumn}% Align table columns on decimal point
\usepackage{bm}% bold math
\usepackage[dvips]{epsfig}
\usepackage{amssymb}
\usepackage{latexsym}
\usepackage{color}
\newcommand \be {\begin{equation}}
\newcommand \ee {\end{equation}}
\newcommand \bea {\begin{eqnarray}}
\newcommand \eea {\end{eqnarray}}

\begin{document}
%\setpagewiselinenumbers
%\modulolinenumbers[5]
%\linenumbers

\title{Pressure independence of granular flow through an aperture}

\author{M. A. Aguirre$^{1}$, J. G. Grande$^{1}$, A. Calvo$^{1}$, L. A. Pugnaloni$^{2}$ and J.-C. G\'{e}minard$^{3}$}
\affiliation{$^{1}$Grupo de Medios Porosos, Fac. de Ingenier\'{\i}a, Universidad de Buenos Aires.
Paseo Col\'{o}n 850, (C1063ACV) Buenos Aires, Argentina.\\
$^{2}$ Instituto de F\'{\i}sica de L\'{\i}quidos y
Sistemas Biol\'{o}gicos (UNLP, CONICET La Plata), casilla de correo 565, 1900 La Plata, Argentina.\\
$^{3}$Universit\'e de Lyon, Laboratoire de Physique, Ecole Normale Sup\'erieure de
Lyon, CNRS, 46 All\'ee d'Italie, 69364 Lyon cedex 07, France.}

\begin{abstract}
We experimentally demonstrate that the flow rate of granular
material through an aperture is controlled by the exit velocity
imposed to the particles and not by the pressure at the base,
contrary to what is often assumed in previous works. This result
is achieved by studying the discharge process of a dense packing
of monosized disks through an orifice. The flow is driven by a
conveyor belt. This two-dimensional horizontal setup allows to
uncouple pressure and velocity and, therefore, to independently
control the velocity at which the disks escape the horizontal
\emph{silo} and the pressure in the vicinity of the aperture.
The flow rate is found to be directly proportional to
the belt velocity, independent of the amount of disks in the
container and, thus, independent of the pressure in the outlet
region. In addition, this specific experimental configuration
makes it possible to get information on the system dynamics from
a single image of the disks that rest on the conveyor belt after
the discharge.
\end{abstract}
%{45.70.-n:Granular systems, 45.70.Mg:Granular flow: mixing, segregation and stratification,
%47.80.Jk:Flow visualization and imaging }
\maketitle
The flow of granular media through an orifice presents
some interesting features that have been intensely studied in the
last $50$ years
\cite{Beverloo,Kadanoff,deGennes,Trappe,Jaeger,Duran,Ristow,Nedderman,Tuzun,Savage,Tighe}.
What mainly differentiates the discharge of a container filled
with granular matter from one filled with a viscous liquid is
that the mass flow-rate does not depend on the height $h$ of
material above the outlet. The explanation most frequently used
for this independence is based on the Janssen's
effect: the distribution of the weight of the material onto the
silo walls, due to the friction forces, leads to a saturation of
the pressure at the bottom, which results in a
constant flow-rate \cite{Reviews}.
In such a reasoning, the pressure is thus
implicitly assumed to govern the flow rate.

Here we show, by using an experimental setup in which
the exit velocity of the grains is decoupled from the bottom
pressure, that the above argument is improper. Indeed, different
flow-rates can be achieved for the same bottom pressure;
moreover, the flow-rate remains constant even if the pressure
decreases during the discharge.

In general, the discharge of a silo through an orifice can
present three regimes: a continuous flow, an intermittent flow,
or a complete blockage of the flow due to arching
\cite{Mankoc07,Mankoc2009, Ulissi}. In the continuous flow
regime, the mass flow rate $W$ is described by the so called
Beverloo's law \cite{Beverloo,BrownBook}: $W= C \rho_{3D}
\sqrt{g}(A-k\,D)^{5/2}$
where $A$ is the size of the opening, $g$ the
acceleration due to gravity, $\rho_{3D}$ the bulk density
and $D$ the diameter of the granules
whereas $k$ and $C$ are two empirical dimensionless constants.
In a two-dimensional (2D) setup---or similarly in slit shaped
apertures---one expects $W= C \rho_{2D}
\sqrt{g}(A-kD)^{3/2}$ \cite{BrownBook}. In the jamming regime, the
jamming probability is controlled by the ratio
$A/D$ of the aperture size to the grain diameter
\cite{Mankoc2009,To,Zuriguel03,Zuriguel05,Janda,Ulissi}.

The discharge of particulate systems is not only found under the
influence of gravity. In many industrial applications, the
grains are horizontally transported at constant velocity
(as on a conveyor belt \cite{desong03} or floating on the surface
of a flowing liquid \cite{Gariguata2009}). Here, we analyze the
discharge through an orifice of a 2D pile of disks driven at a
constant velocity by a horizontal conveyor belt.

\begin{figure}[!t]
\includegraphics[height=\columnwidth,angle=-90]{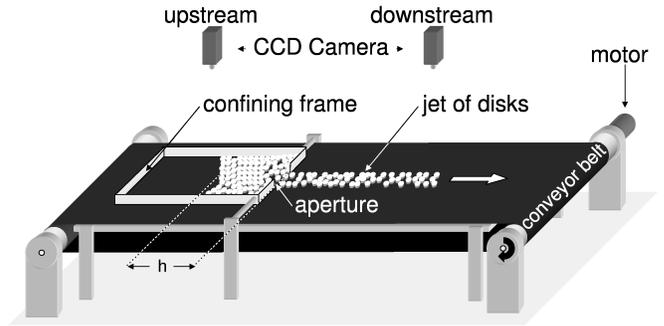}
\caption {Sketch of the experimental setup- Disks, placed on a
conveyor belt which moves at a constant velocity $V$ (white
arrow), are forced to flow through an aperture in the confining
frame. Observation of the grains---either remaining in the
confining frame or forming the {\it jet} outside---makes it
possible to study the dynamical properties of the discharge.}
\label{ExpSetup}
\end{figure}

The experimental setup (Fig.~\ref{ExpSetup}) consists of a conveyor
belt (width $40$~cm, length $1$~m) above which a confining
Plexiglas frame (width 26~cm, length 54~cm) is maintained at a
fixed position in the frame of the laboratory. A motor drives the
belt at a constant velocity $V$ which ranges from $0.4$~cm/s to
$4$~cm/s. Downstream, the Plexiglas frame exhibits, at the center,
a sharp aperture (width $A \le 10$~cm), much wider
than the thickness of the walls ($0.6$~cm).
The granular material consists of $N = 450$ Plexiglas disks
[thickness $e = (3.1 \pm 0.1)$~mm, and diameter $D =
(1.04 \pm 0.01)$~cm].
Thus, in our experimental conditions, the ratio $A/D \le 10$.

Before the flow is started, the initial state of the system is
obtained by depositing, in a disordered manner, $N$ disks on the
conveyor belt inside the confining frame. The belt is then moved
at a low constant velocity until all grains are packed against the
downstream wall, whose aperture is still kept closed ($A = 0$).
The packing fraction of this initial configuration is
$\Phi\approx0.82$. Then, the aperture is opened to the
desired width $A \neq 0$ and the belt is moved at 
constant velocity $V$. The belt is stopped whenever the Plexiglas
frame is emptied or the flow ceases due to the formation
of a jamming arch.

A video camera (Pixelink, PL-A741) is used to image the system from above. We record the evolution of
the discharge process in either of the following ways: Images of disks that remain inside the confining
frame (upstream) are acquired every second during the discharge; One single image of the \textit{jet}
of disks that are at rest on the belt outside the confining box (downstream) is acquired after the discharge.

\begin{figure}
\includegraphics[width=\columnwidth]{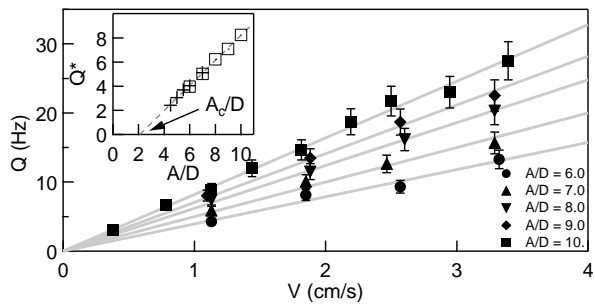}
\caption {Dependence of the flow rate $Q$ on the belt velocity
$V$ for different apertures $A/D$ (Average of 4 runs).
Inset: The dimensionless slope
$Q^*\equiv QD/V$ obtained fitting data reported in the main panel
(open squares) and/or from the probability distributions $P(Q^*)$
(crosses) reported in Fig.~\ref{distributions} (errors are $\pm
5\%$). The flow rate is expected to vanish for the critical flow
rate, $A_c$.} \label{caudal}
\end{figure}
An intensity threshold is used to convert the pictures into
binary images. 
The number $N_i$ of disks that remain inside the confining frame at
time $t$ can be calculated from the number of white pixels in the
images acquired upstream [Area of a disk ($733 \pm 8$) pixels].
The instantaneous disk flow-rate (averaged over 1 s, because of the acquisition frequency) is
defined as $Q \equiv -\frac{dN_i}{dt}$. When the aperture is
large ($A/D \ge 6$), the flow-rate is rather continuous
throughout the discharge, $N_i$ depends linearly on
time and, for any given $A$, the flow rate is
proportional to the belt velocity (Fig.~\ref{caudal}).
This indicates that, in our experimental conditions, the granular bed
rearranges in a time which is smaller or compares with
the time necessary for the grains to escape the outlet region.
Indeed, if a significant relaxation time is
associated with the rearrangement of the granular bed,
above a critical velocity, a change in the
drive is barely followed by the system and a qualitative
change in $Q(V)$ is observed \cite{desong03}.
In the inset in Fig.~\ref{caudal}, we report the dimensionless slope
$Q^*\equiv QD/V$ as a function of the dimensionless aperture
$A/D$ and get, for $A/D \ge 6$, the empirical law: 
\begin{equation}
Q = {\cal C} \frac{V}{D} \Bigl(A/D - k\Bigr)
\label{beverloo}
\end{equation}
where $k = 2.1 \pm 0.2$ and ${\cal C} = 1.04 \pm
0.02$. The flow rate is thus expected to vanish at
a value $A_c \equiv k D$ of the aperture.
The value of the constants $k$ and $\cal C$ will be
discussed below (see comment on Ref.~\cite{desong03}). 
Note that the flow rate is not well defined and cannot be
extracted directly from $N_i(t)$ for $A/D < 6$, the
system being then likely to jam.

\begin{figure}
\includegraphics[width=\columnwidth]{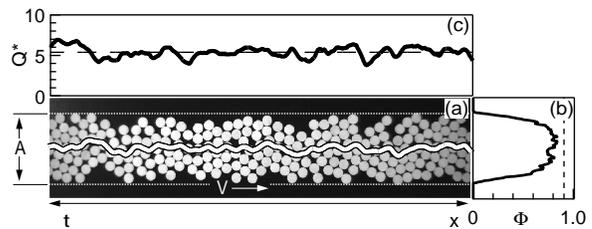}
\caption {Image of the {\it jet} and the corresponding density
profiles. (a) Image of the jet after the discharge of the
container for $A/D = 7.0$. (b) Area fraction profile $\Phi$ in
the perpendicular direction. Each data point is obtained as the
fraction of white pixels in the corresponding row after
thresholding [the vertical dashed line corresponds to the maximum
attainable density $\pi/(2\sqrt{3})$]. (c) Instantaneous
dimensionless flow rate $Q^*$ against the longitudinal coordinate
$x$. The flow rate is obtained from the fraction of white pixels
in the corresponding column $1D$ wide.} \label{foto}
\end{figure}

In order to estimate the flow rate for $A/D<6$, it is
particularly interesting to consider the second measurement
method in which we take a single picture of the jet at rest on the
belt after the experiment is stopped (Fig.~\ref{foto}a).
After being released from the confining frame, the grains are simply
advected with velocity $V$ and the belt is long enough so that all
disks remain on it after the discharge. Thus, the distribution of
disks on the belt provides information on the outflow of grains
ever since its onset. Therefore, such an image provides the full
history of the flow rate: the spatial coordinate along the belt
longitudinal axis can be translated into time. Moreover, since
the flow rate $Q$ is proportional to $V$ (Fig.~\ref{caudal}), one
can limit the study to a single value of the velocity.

Let us first comment on the intensity profile in the direction
perpendicular to the jet (Fig.~\ref{foto}b): the fraction of
white pixels in a row (one pixel wide and about $1500$ pixels
long) corresponds to the average density $\Phi$ of the disks in
this row. The profile is smooth and, even for an aperture as
large as seven disk diameters, there is no well defined plateau.
Such boundary effects, which are present in a large part of the
profile, are responsible for the effective aperture width, $(A -
A_c)$, pointed out by the finite value of $k$ in
Eq.~(\ref{beverloo}).

\begin{figure}
\includegraphics[width=\columnwidth]{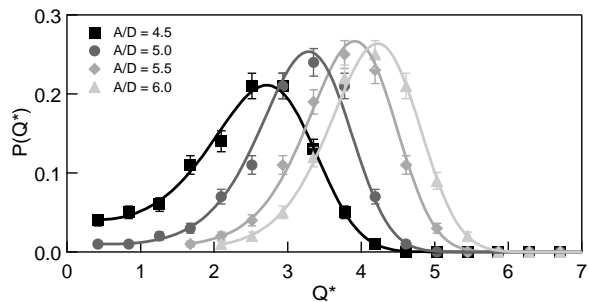}
\caption {Probability distribution of the dimensionless flow rate
$Q^*$ for different values of $A/D$. For each aperture $A/D$, the
probability is calculated from all the values of the flow-rate
measured during a set of 5 to 10 experiments. For the largest
aperture reported ($A/D = 6.0$), the probability distribution is
close to Gaussian, whereas a relatively large probability of
small $Q^*$ values appears when $A$ is decreased. The continuous
lines are only a guide to the eye.} \label{distributions}
\end{figure}

We now consider the intensity profile along the longitudinal axis
of the jet. In order to define an instantaneous flow rate, we
arbitrarily integrate over a 30-pixel-wide window (a disk
diameter) around the considered position. The dimensionless
flow rate $Q^*$ is obtained by counting the number of white
pixels in a column of width $D$ along the jet and dividing it
by the surface area of one disk. Other window widths were tested:
fluctuations are smaller for wider windows but the average value
of $Q^*$ remains the same to within 0.2\%. The example
reported in Fig.~\ref{foto}c shows that, even for a large
aperture, $Q^*$ is subjected to significant fluctuations
(standard deviation around $15$\% and difference between extreme
values of about $60$\% of the average). The probability
distribution $P(Q^*)$ is shown in Fig.~\ref{distributions}.
On the one hand, $P$ is almost Gaussian when the aperture is large ($A/D>6$).
The probability of 
small $Q^*$ is negligible and the flow is continuous. On the
other hand, when the size of the aperture is reduced, the system
might jam ($Q^*=0$) transitorily or permanently (in agreement
with similar experiments carried out in a 2D vertical silo
\cite{Janda2009, Ulissi}). In accordance, the
probability of small $Q^*$ increases.

From the probability distribution, it is possible to define a
mean flow rate even when the system presents intermittent jams.
The average flow rate during the effective flow events can be
estimated by defining $\bar{Q}^*\equiv\int Q^*P(Q^*)dQ^*$ (Note
that jamming events correspond to a peak at $Q^*=0$ that we
disregard in our analysis in order to account for the flow events
only). The estimations of $Q^*$ obtained in this way for large
values of $A/D$ are in full agreement with measurements obtained
from the upstream technique (Fig.~\ref{caudal}, inset). In
addition, the data obtained for smaller apertures also obey
to Eq.~(\ref{beverloo}), which indicates that there is no drastic
change in the empirical law in presence of intermittency.
However, we mention that, in $3D$ configurations, a
deviation from the $5/2$
Beverloo's scaling for very small apertures has been observed
\cite{Mankoc07}. Thus, a deviation from the linear relation
(\ref{beverloo}) may be present for $A/D < 4.5 $.

We emphasize that, in our experimental conditions, the effective
pressure exerted by the granular bed on the downstream wall
depends on the dynamical friction force between the disks and the
belt. So, at a constant velocity $V$, the bottom pressure continuously
decreases as the number of disks inside the system decreases
during the discharge.
Thus, during the whole discharge, a constant flow-rate $Q$
is achieved while the pressure continuously varies, which indicates
that the granular flow rate is not controlled by it.
The horizontal configuration makes it possible to impose the exit velocity and,
moreover, to tune independently the overall pressure by changing the weight of the disks
or their initial number inside the frame. In order to measure the
flow-rate in a different pressure range for the same output
velocity, we added an individual extra weight on top of each
disk. We thus increased the pressure by increasing the friction
force acting on the disks without changing any other property of
the system. We did not observe any change in the flow-rate in
spite of the change in the pressure, which is again in agreement
with a flow-rate only controlled by the escape velocity.

At this point, it is particularly interesting to discuss the
physical ingredients leading to Beverloo's law. In our
experimental conditions, we obtain a simple linear law relating
the flow-rate to the aperture width $A$, whereas the corresponding
relation in a 2D gravity driven configuration corresponds to $Q
\propto A^{\frac{3}{2}}$. Both relations can be derived from a
dimensional analysis, taking into account the estimation of the
velocity $v$ at which the grains escape the system. In the case
of the 2D vertical silo, the \textit{free falling arch} assumption
\cite{NeddermanBook} leads to $v \propto \sqrt{g A}$. Hence,
considering the mass flow rate through an aperture of size $A$,
$W \equiv \rho_{2D} A v$ leads to $W \propto \rho_{2D} \sqrt{g}
A^{\frac{3}{2}}$. Boundary effects at the aperture edges lead to
two boundary layers of thickness $\sim D$ (the so-called
\textit{empty annulus} \cite{Brown1961}). Considering the effective
aperture size $A - k D$ instead of $A$, one gets $Q \propto (A/D
- k)^{\frac{3}{2}}$. The same arguments lead to the Beverloo's
law in 3D. In the case of the conveyor belt $v=V$ and the account
of the empty annulus leads to $Q \propto V (A/D - k)$ in
agreement with Eq.~(\ref{beverloo}). It is then worth mentioning
that, in the limit of infinitely large apertures, the correction
due to the boundary layers can be neglected and $Q \to \rho A V$
($\rho$ is the number of disks per unit area) so that $C = \rho
D^2$ and, hence, $Q = \rho V D (A/D - k)$. Taking into account
$\Phi = \rho \frac{\pi D^2}{4}$ ($\Phi$ is the area
fraction), we obtain that ${\cal C} = \frac{4}{\pi}\,\Phi$. From
the experimental value ${\cal C} = 1.13 \pm 0.02$, we estimate
$\Phi = 0.82 \pm 0.02$, which is indeed compatible with the area
fraction initially measured inside the confining frame and which
does not noticeably vary during the discharge. We remark that the
dimensional analysis could be based on the pressure since it also
exhibits a time-scale. In the case of a 3D silo, denoting $P$ the
pressure at the outlet, one gets $W \propto \sqrt{\rho_{3D}\,P}
A^2$. Be the pressure limited by Janssen's effect or not, $W$
scales with $A^2$, which contradicts the experimental $A^{5/2}$
scaling.

In most previous works \cite{Reviews}, the independence of the
flow rate on the silo height $h$ is explained by means of two
postulates. The first one is: if the height is larger than twice
the diameter of the silo, the pressure at the bottom is constant
(Janssen's effect). The second one is: if during the discharge
the pressure at the bottom is constant, the flow rate is, as a
consequence, also constant. We have proven that this second
postulate is not valid. Indeed, experimentally, for the same
pressure in the outlet region (same amount of disks inside the
system), the flow rate is proportional to the belt velocity.
Moreover, the same flow-rate is obtained with different values of
the pressure in the outlet region (different amount of disks
inside the system or additional weights on the disks). In
addition, theoretically, the physical ingredients used to
establish Beverloo's law, which is correctly satisfied
experimentally, do not involve the local pressure but a
characteristic velocity. Thus, Beverloo's law already predicts
that the flow-rate should be independent of $h$ even when the
Jannsen's effect is not at stake.

In conclusion, using an experimental setup which enables to
control the particle velocity independently from the other
parameters of the flow, we have shown that the granular flow rate
through an orifice is not controlled by the local pressure and
that invoking the Jannsen's effect is not pertinent to explain the
constant flow-rate measured during a silo discharge. The flow
rate is controlled by the mechanism driving the grains out of the
system and by the geometry of the outlet, features which govern
the exit velocity, and not by the pressure upstream. For
applications, our findings thus suggest that granular flow rates
can be increased, in gravity driven systems, by locally
increasing the particle velocities at the outlet, which might be
achieved with devices having a low energetic cost. From the
fundamental point of view, they show that more effort must
certainly be done to verify some misleading assertions widely
stated in material science of dissipative particles during the
last five decades.

\acknowledgments We thank J.P. Hulin for valuables comments. This
work has been supported by the program PICT $32888$ (ANPCyT) and
the International Cooperation Program CONICET-CNRS. J.G.G.
acknowledges support from UBA, L.A.P. and M.A.A. from CONICET.

\end{document}